\documentclass[3p,times,procedia]{elsarticle}
\usepackage{nupha_ecrc}

\volume{00}

\firstpage{1}

\journalname{Nuclear Physics A}

\runauth{}

\jid{nupha}

\jnltitlelogo{Nuclear Physics A}

\usepackage{amssymb}

\usepackage[figuresright]{rotating}

\newcommand{\sNN}{s_\mathrm{NN}}
\newcommand{\dd}{\partial}

\begin{document}

\begin{frontmatter}



\dochead{XXVIIth International Conference on Ultrarelativistic Nucleus-Nucleus Collisions\\ (Quark Matter 2018)}

\title{Lambda polarization in heavy ion collisions: from RHIC BES to LHC energies}


\author[IK]{Iurii Karpenko}
\author[FB]{Francesco Becattini}

\address[IK]{SUBATECH, IN2P3/CNRS, %
  Universit\'{e} de Nantes, IMT Atlantique, %
  4 rue Alfred Kastler, 44307 Nantes cedex 3, %
  France}
\address[FB]{
Universit\'a di Firenze and INFN - Sezione di Firenze, Via G. Sansone 1, I-50019 Sesto Fiorentino (FI), Italy
}
\begin{abstract}
STAR collaboration at RHIC has recently measured the polarization of $\Lambda$ hyperons in non-central heavy ion collisions in the RHIC Beam Energy Scan (BES) program. The magnitude of the polarization was found to decrease from few percents at the lowest BES energies to $\approx0.2$\% at the top RHIC energy. The polarization signal has been reproduced in different hydrodynamic calculations assuming a thermodynamic spin-vorticity coupling mechanism at the Cooper-Frye hypersurface \cite{Becattini:2013fla}.

In this work an extension of our existing calculations \cite{Karpenko:2016jyx} of the $\Lambda$ polarization in the RHIC BES program to the top RHIC and 2.76~TeV LHC energies is presented. The longitudinal component of the $\Lambda$ polarization, which is the dominant component of the polarization at the LHC energies, is discussed. Finally we show that the global polarization of $\Lambda$ originates dominantly from the relativistic analogue of the classical vorticity, whereas the quadrupole longitudinal component originates from the gradients of temperature and acceleration of the medium when the $\Lambda$s are produced out of the fluid.
\end{abstract}


\end{frontmatter}



\section{Introduction}
\label{intro}
STAR collaboration has recently discovered a significantly nonzero global polarization of $\Lambda$($\bar\Lambda$) hyperons produced in non-central Au-Au collisions in the RHIC Beam Energy Scan (BES) Program \cite{STAR:2017ckg}. Following this discovery, an extended analysis has been performed at the top RHIC energy \cite{Adam:2018ivw}, and found the polarization signal to be in line with the trend from the BES energies.
Different hydrodynamic models \cite{Karpenko:2016jyx,Xie:2017upb} and AMPT parton/hadron cascade \cite{Li:2017slc, Sun:2017xhx} generally reproduce the magnitude of the measured polarization. In the hydrodynamic models the $\Lambda$ hyperons produced at particlization (fluid to particle transition) hypersurface acquire polarization via a thermodynamic spin-vorticity coupling mechanism. In the AMPT calculations the same spin-vorticity coupling mechanism is introduced with the help of a coarse-graining procedure. The magnitude of polarization depends on local {\it thermal vorticity} of the fluid, $\varpi_{\mu\nu}=(\dd_\nu \beta_\mu - \dd_\mu \beta_\nu)$ at the points of $\Lambda$ production, $\beta_\mu=u_\mu/T$ being the inverse four-temperature field.

The global (i.e.\ $p_T$ integrated) polarization of the $\Lambda$($\bar\Lambda$) hyperons is directed perpendicular to the reaction plane and parallel to the vector of the global angular momentum of the system. The magnitude of the global polarization is found to decrease fast from lowest to highest RHIC BES energies, both in the experiment and in the theory calculations. In this proceeding we report on the extension of the $\Lambda$ polarization calculations to 2760~GeV LHC energy and discuss the component of polarization along the beam direction.

\section{Results}
\label{sec-results}
The global polarization of $\Lambda$ hyperons in heavy ion collisions at the RHIC Beam Energy Scan energies $\sqrt{\sNN}=7.7\dots 200$~GeV in the framework of a 3 dimensional viscous hydrodynamic model vHLLE+UrQMD has been reported in \cite{Karpenko:2016jyx}. Now we have extended the calculation to the 2.76 TeV LHC energy. For this purpose, instead of previously used initial state (IS) from the UrQMD cascade, we have employed a Monte Carlo Glauber IS with parametrized rapidity dependence as in \cite{Bozek:2012fw}. The initial angular momentum of the system in this initial state model is induced by a tilted initial energy density profile in the $x-\eta$ plane. Such tilted initial energy density profile is consistent with a preferential emission in the forward-backward hemisphere from forward-backward participating nucleons, and has been used to study the directed flow in heavy ion collisions at $\sqrt{\sNN}=200$~GeV \cite{Bozek:2010bi}.

\begin{figure}
\includegraphics[width=0.5\textwidth]{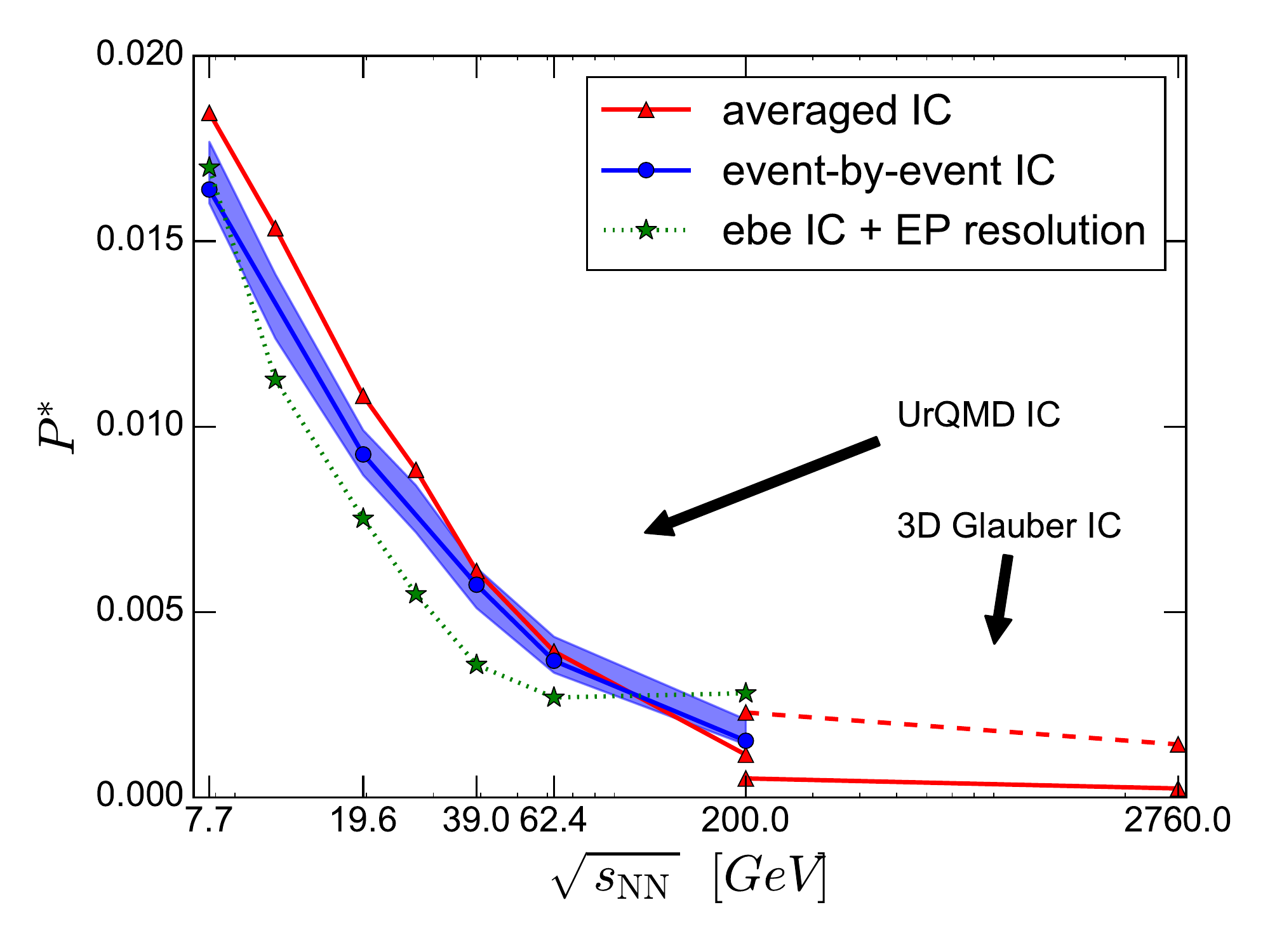}
\includegraphics[width=0.5\textwidth]{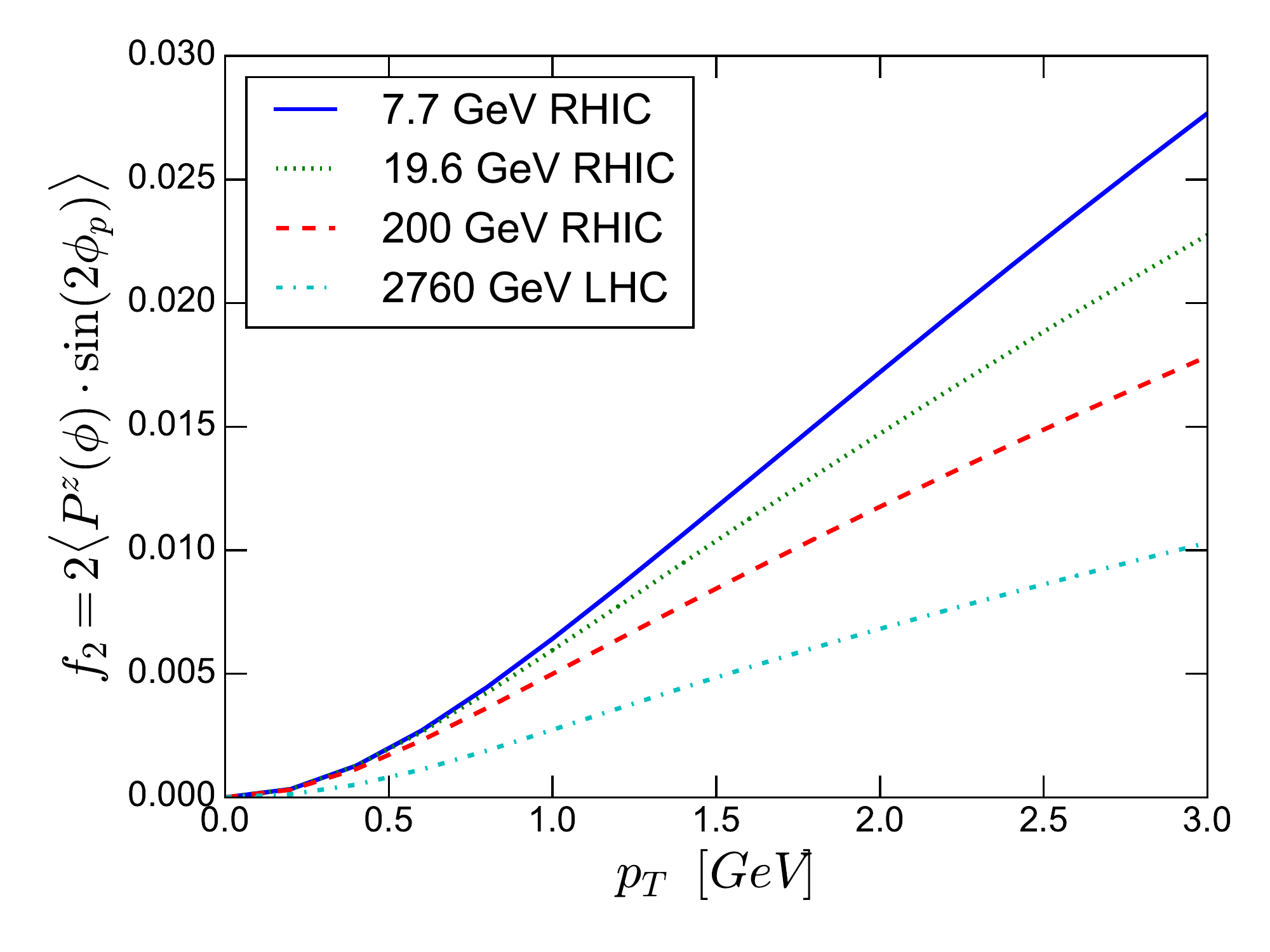}
\caption{Left panel: Global polarization of $\Lambda$ hyperons in 20-50\% central Au-Au (Pb-Pb) collisions at 7.7\dots 200~GeV RHIC (2760 GeV LHC) energies. For the calculations with 3D Glauber IC (initial conditions), the solid one corresponding to longitudinally boost invariant initial flow, and the dashed one corresponding to a small amount of initial shear longitudinal flow as described in \cite{Becattini:2015ska}. The lines connect the points to guide the eye.
Right panel: Second order Fourier harmonic coefficient of polarization component along the beam direction, calculated as a function of $p_T$ for different collision energies; 200 and 2760 GeV points correspond to Monte Carlo Glauber IS.}\label{fig-LHC}
\end{figure}

We observe that, in line with the results from the RHIC BES energies, the global polarization decreases further down from top RHIC to 2.76 TeV LHC energy, see Fig.~\ref{fig-LHC} left. The decrease is found in both calculations with boost-invariant initial flow (solid line connecting 200 and 2760 GeV points) and with small amount of an added shear flow (dashed line). The reasons are slightly lower initial vorticity and a longer lifetime of the hydrodynamic stage at the LHC energy. Therefore the global polarization of $\Lambda$ hyperons at midrapidity in the non-central heavy ion collisions at the LHC is expected to be small ($\le$0.1\%).

The early calculations for the top RHIC energy \cite{Becattini:2015ska} have revealed that in the $p_T$ differential analysis all components of the polarization vector are non-vanishing. In fact a component which has the largest amplitude in the transverse momentum ($p_x p_y$) plane is the one in the beam direction, $P^z$. It has a quadrupole structure in the transverse momentum plane, and vanishes at $p_x=0$ or $p_y=0$. Therefore it is natural to decompose it into the Fourier series, where only sine terms of even multiples of the azimuthal angle $\phi$ of the transverse momentum vector are non-vanishing:
\begin{equation}\label{szfour}
 P^z({\bf p}_T,Y=0) = \sum_{k=1}^\infty f_{2k}(p_T) \sin 2 k \varphi.
\end{equation}
It was later found that the $P^z$ is related to the anisotropy of the transverse expansion \cite{Becattini:2017gcx}. As such, the $P^z$ does not vanish in the case of perfect boost invariance, and can be accessed with a 2D hydrodynamic calculation for non-central heavy ion collision.
Indeed, one can do a simple exercise and consider a Blast-Wave model with an anizotropic transverse momentum distribution which results in a finite elliptic flow coefficient $v_2$ of produced hadrons. Assuming that at the Bjorken proper time of the particle emission $\tau$ there is a time gradient $dT/d\tau$ of the temperature $T$ (which in Bjorken picture depends only on proper time $\tau$), one can show that there is a linear relation between the $p_T$ dependent $f_2$ harmonic and elliptic flow:
\begin{equation}\label{eq-BW}
   f_2(p_T) = 2 \frac{d T}{d \tau}\frac{1}{mT} v_2(p_T)
\end{equation}
where $m$ is mass of the particle.

As one can see from the realistic 3 dimensional viscous hydrodynamic calculations with Monte Carlo Glauber IS on Fig.~\ref{fig-LHC} right, $f_2$ decreases with collision energy much slower than the global polarization. Whereas the global polarization (Fig.~\ref{fig-LHC} left) decreases by about a factor 10 between $\sqrt{\sNN}=7.7\ {\rm and}\ 200$~GeV, $f_2$ decreases by only 35\%. A qualitative explanation for the energy dependence of $f_2$ comes from the Eq.~\ref{eq-BW}: whereas the elliptic flow slightly grows with collision energy, the gradients at the end of hydrodynamic stage decrease due to the longer expansion time.

The origins of the different components of $\Lambda$ polarization can be seen as follows. The polarization of $\Lambda$ hyperons produced at the particlization hypersurface, is up to the Cooper-Frye factor proportional to the vector product of the thermal vorticity tensor and 4-momentum vector. Because the thermal vorticity $\varpi_{\mu\nu}=(\dd_\nu \beta_\mu - \dd_\mu \beta_\nu)$ is expressed in terms of the inverse temperature field $\beta_\mu=u^\mu/T$, its vector product with the 4-momentum of the particle can be therefore decomposed into contributions from gradient of temperature, relativistic extension of non-relativistic vorticity and relativistic acceleration, as follows:

\begin{equation}\label{eq-dcmp}
S^\mu\propto \epsilon^{\mu\rho\sigma\tau}\varpi_{\rho\sigma}p_\tau=\epsilon^{\mu\rho\sigma\tau}(\dd_\rho \beta_\sigma)p_\tau=
\underbrace{\epsilon^{\mu\rho\sigma\tau}p_\tau\dd_\rho\left(\frac 1 T\right)u_\sigma}_{{\rm grad} T}\ +\ 
\underbrace{\frac{1}{T}2\left[\omega^\mu(u\cdot p)-u^\mu(\omega\cdot p)\right]}_{\rm vorticity}\ +\ 
\underbrace{\frac{1}{T}\epsilon^{\mu\rho\sigma\tau}p_\tau A_\sigma u_\rho}_{\rm acceleration}
\end{equation}
where the acceleration field is $A_\sigma=u^\lambda \dd_\lambda u_\sigma$ and $\omega^\mu={1\over2} \epsilon^{\mu\rho\sigma\tau}\omega_{\rho\sigma}u_\tau$ is a relativistic extension of the angular
velocity pseudo-vector, which is expressed in terms of kinematical vorticity $\omega_{\rho\sigma}$.

\begin{figure}
\includegraphics[width=0.5\textwidth]{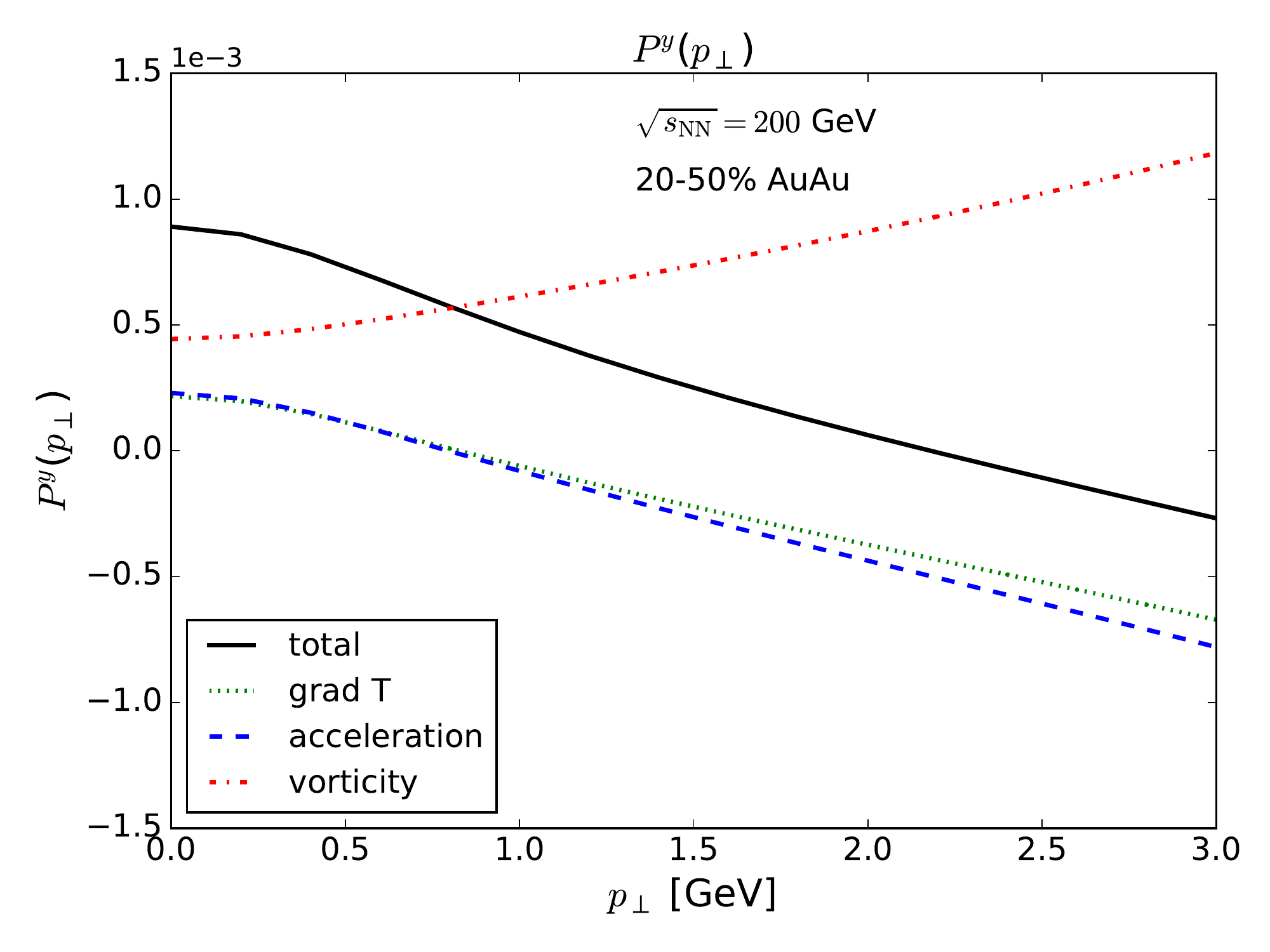}
\includegraphics[width=0.5\textwidth]{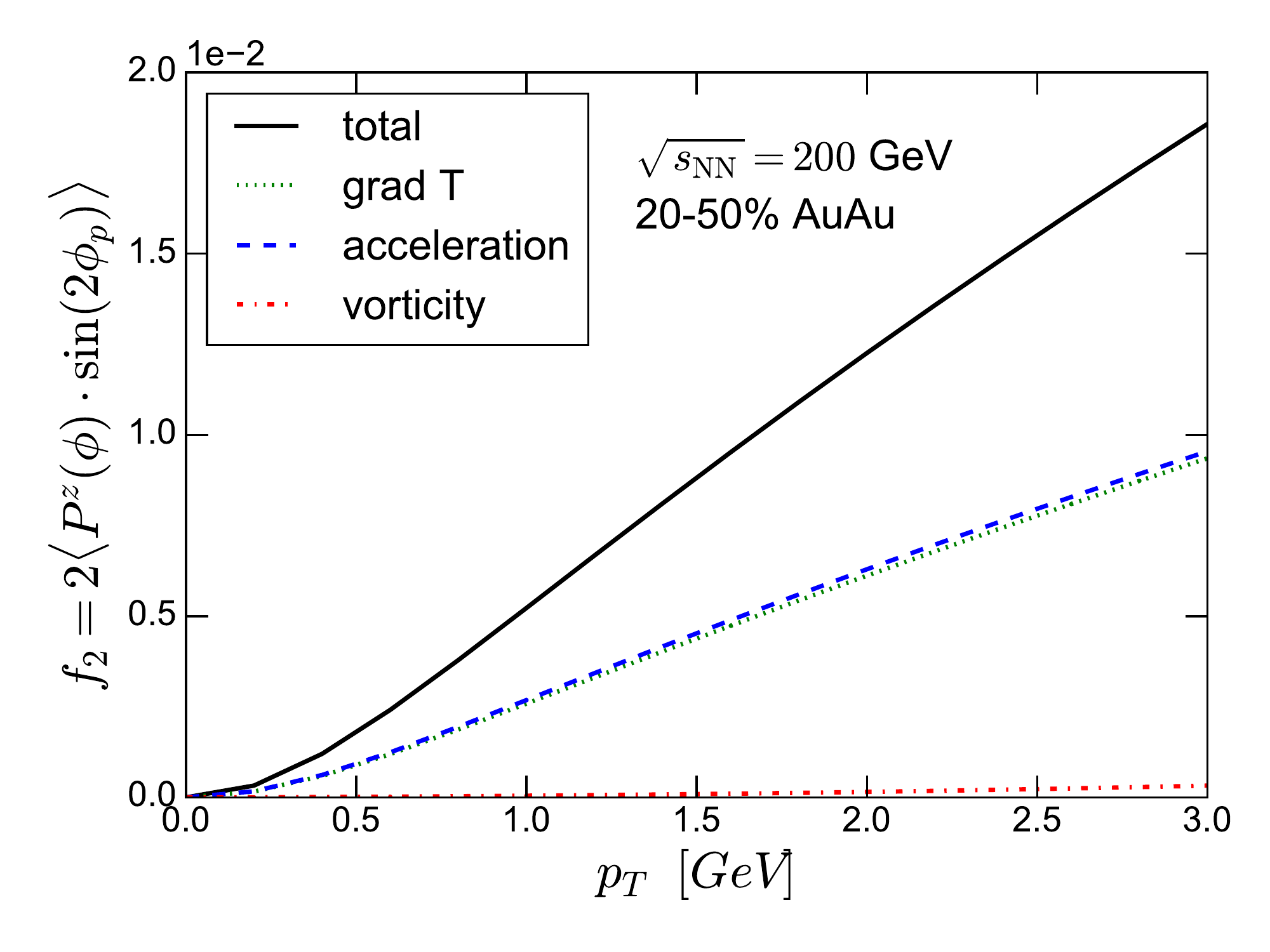}
\caption{Contributions to the global (left panel) and quadrupole longitudinal (right panel) components of $\Lambda$ polarization stemming from gradients of temperature (dotted lines), acceleration (dashed lines) and vorticity (dash-dotted lines). Solid lines show the sums of all 3 contributions. The calculations are done with averaged MC Glauber IS corresponding to 20-50\% central Au-Au collisions at 200~GeV RHIC energy.}\label{fig-contrib}
\end{figure}

On Fig.~\ref{fig-contrib} we plot the contributions to the global and quadrupole longitudinal polarization components from gradients of temperature, acceleration and vorticity individually, as well as their sum. One can see that the resulting $p_T$-integrated global polarization of $\Lambda$, which is dominated by its low-$p_T$ contributions, has the largest contribution from the classical vorticity term. At the same time, $f_2$ has a negligible contribution from the vorticity term and virtually equal contributions from the grad T and acceleration terms. The latter result is expectable, as in hydrodynamics of ideal uncharged fluid the temperature gradient and acceleration fields are related as follows:
\begin{equation}
 A_\mu = \frac{1}{T}\Delta_{\mu\nu}\dd^\nu T
\end{equation}
Therefore in the case of ideal uncharged fluid the grad T and acceleration contributions will be exactly equal to each other. Thus the small difference between the grad T and acceleration contributions seen on Fig.~\ref{fig-contrib} shows that, even though the shear viscosity over entropy ratio in the calculations changes between $\eta/s=0.08\dots 0.2$, the resulting hydrodynamic evolution is quantitatively not very different from ideal one.

A preliminary result from the STAR experiment for the $\Lambda$ polarization component in the beam direction in 10-60\% Au-Au collisions at $\sqrt{\sNN}=200$~GeV RHIC energy has been presented at this conference \cite{Takafumi-QM2018}. The azimuthal angle dependence of the polarization has the same $\sin(2\phi)$ dependence, however its amplitude is smaller and has an opposite sign to the calculations in hydrodynamic models. Such discrepancy between the experiment and theory calculations remains an open questions for further studies.

\section{Acknowledgements} This work was partly supported by the University of Florence
grant {\it Fisica dei plasmi relativistici: teoria e applicazioni moderne}. IK acknowledges support by Region Pays de la Loire (France) under contract no.~2015-08473.





\bibliographystyle{elsarticle-num}
\bibliography{refs}







\end{document}